\documentclass[12pt]{iopart}

\usepackage{graphicx}
\usepackage{xcolor}
\usepackage{amssymb}
\usepackage{setstack}
\usepackage[normalem]{ulem}
\usepackage{soul}
\newcommand{\green}[1]{{\color[rgb]{0,0,0}{#1}}}

\begin{document}

\title{Non-classicality of coherent state mixtures}

\author{I. Starshynov, J. Bertolotti, J. Anders}

\address{Department of Physics and Astronomy, CEMPS, University of Exeter, Stocker Road, Exeter EX4 4QL, UK}
\ead{is283@exeter.ac.uk, \, janet@qipc.org}

\begin{abstract}

Mixtures of coherent states are commonly regarded as classical. 
Here we show that there is a quantum advantage in discriminating between coherent states in a mixture, implying the presence of quantum properties in the mixture, which are however not captured by commonly used non-classicality measures. 
We identify a set of desired properties for any non-classicality measure that aims to capture these quantum features, and define the discord potential $C_D$, which we show to satisfy all those properties. 
We compare the discord potential with recently proposed coherence monotones, and prove that the coherence monotones diverge for classically distinguishable states, thus indicating their failure to quantify non-classicality in this limit. 
On the technical side, we provide a simple method of calculating the discord as well as other information-theoretic quantities for the (non-Gaussian) output of any input state with positive P-function.
\end{abstract}

\color{black}

%%%
\section{Introduction} \label{intro}
It is often not obvious whether a given state of a physical system requires a quantum-mechanical description, or whether a classical description suffices. Moreover, on closer inspection one realises that a physical system can show distinct quantum features~\cite{Ferraro2012}, such as non-positive ``probabilities'', and correlations that violate Bell's inequality. Hence different measures of  ``quantumness'' have been introduced that each quantifies how much a quantum state exhibits a particular feature~\cite{Dodonov,Ryl2015,Yuan2018,Bernardini2017}. Popular non-classicality measures include ill-definiteness or negativity of the Glauber-Sudarshan P-function~\cite{Gilchrist1997,Glauber1963}, the entanglement potential~\cite{Asboth2005}, and the recently proposed coherence monotones~\cite{Baumgratz2014,Streltsov2016}.

An interesting case study are proper mixtures of coherent states, $\sum_j p_j \, \left| \alpha_j \right \rangle \left \langle \alpha_j \right|$, where $p_j$ are positive probabilities, and $\left| \alpha_j \right \rangle$ are coherent states~\cite{Loudon2000}. Because coherent states are commonly thought to be classical, and mixing is also a classical operation, one expects mixtures of coherent states to be classical, too. And indeed their P-function is well defined and positive, and they have zero entanglement potential.  However, since any two coherent states are never completely orthogonal there is always an ambiguity in discriminating them from each other. It is well-known that quantum measurements are able to discriminate coherent states in a mixture better than any classical measurement~\cite{Helstrom1967,Barnett2009}, suggesting that mixtures of coherent states do have a ``quantumness'' that is not quantified by neither the P-function nor the entanglement potential~\cite{Hosseini2014,Choi2017}.

In this paper we propose a new measure of non-classicality, the discord potential $C_D$, that captures this feature. In Section~\ref{sec:quadvantage} we discuss the quantum advantage in state discrimination. In Section~\ref{sec:discordpot}  we identify the properties that we look for in a non-classicality measure and define the discord potential, which satisfies all the desired properties. In Section~\ref{sec:properties} we discuss the properties of the discord potential for mixtures of two coherent states. We also show that while two recently proposed \emph{coherence monotones} diverge even when the quantum advantage in state discrimination disappears, the discord potential correctly goes to zero.  A formal proof of the divergence of two coherence monotones is presented in the Appendix (Section~\ref{sec:app}). We provide a method to calculate entropies and discord potential for any proper mixture of coherent states (including non-Gaussian states) in Section~\ref{sec:method}, and summarise our results 

in Section~\ref{sec:concl}.

%%%
\section{Quantum advantage in state discrimination} \label{sec:quadvantage}

State discrimination is the problem of identifying which of a set of expected states has been received. If the states are orthogonal this is a trivial task, however, if the possible states are non-orthogonal the discrimination will always have an associated error. This error will depend on what measurements are performed on the system, and proving optimal state discrimination is an area of active research~\cite{Wittmann2008,Cook2007,Becerra2013,Rosati2016}.

\begin{figure}[!t]
\centering
    \includegraphics[scale=1]{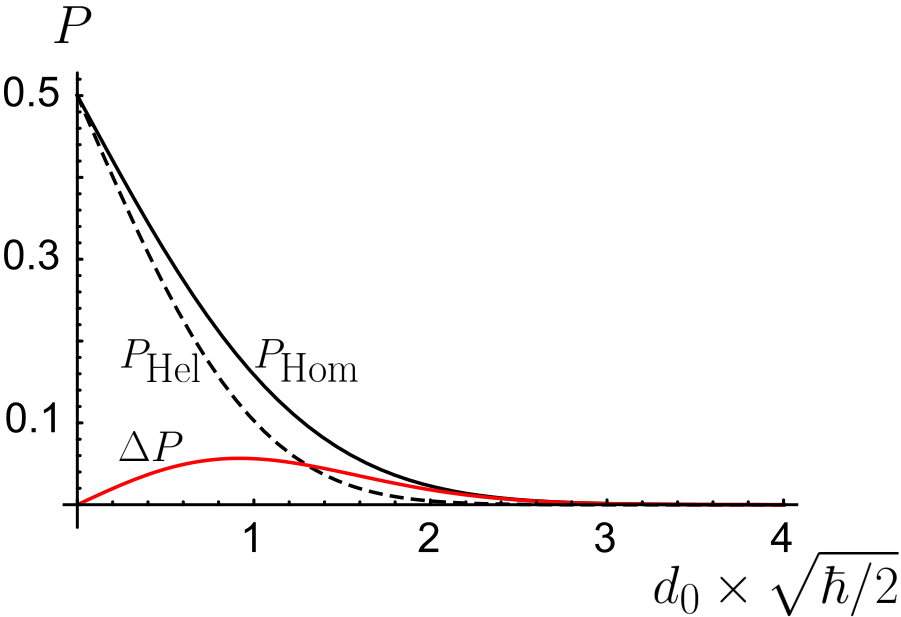}
    \caption{ \label{fig:Helstrom} Probability $P$ of incorrectly discriminating between $| \alpha_0 \rangle $ and $| \beta_0 \rangle$ as a function of their separation $d_0=| \alpha_0  - \beta_0|$ and for $a=1/2$.  Error probability for a homodyne measurement, $P_{\mathrm{Hom}}$ (black solid line),  optimal error probability given by the Helstrom bound, $P_{\mathrm{Hel}}$ (black dashed line), and their difference $\Delta P = P_{\mathrm{Hom}}- P_{\mathrm{Hel}}$ (red solid line). }
\end{figure}

Let us consider the simple example of distinguishing two coherent states $|\alpha_0 \rangle$ and $|\beta_0 \rangle$ that occur with probabilities $0 < a < 1$ and $1- a$, respectively, i.e. the corresponding mixed state is
\begin{equation}\label{eq:rho}
	\rho_0  = a \, \vert \alpha_0 \rangle  \langle  \alpha_0 \vert + (1-a) \, \vert \beta_0 \rangle \langle \beta_0 \vert .
\end{equation}
When the separation $d_0$, which is related to the states' overlap  $d_0^2:= |\alpha_0 - \beta_0|^2 = - \ln |\langle \alpha_0 \vert \beta_0 \rangle|^2$, decreases to zero the two coherent states become identical and the system is not a mixture anymore. For large separations, the two coherent states have less and less overlap and thus become classically distinguishable. But for intermediate separations there will be a non-trivial probability $P(a,d_0)$ of identifying a state incorrectly, which will have a maximum for $a=1/2$. Interestingly,  when discriminating between $| \alpha_0 \rangle $ and $| \beta_0 \rangle$, quantum measurements have an advantage over classical (homodyne) ones. I.e. the optimum measurement produces an error  given by the Helstrom bound~\cite{Helstrom1967}, $P_{\mathrm{Hel}} (1/2,d_0)~=~\frac{1}{2}\left[1-\sqrt{1-e^{-d_0^2}}\right]$,   which is less than the corresponding error for homodyne measurements~\cite{Wittmann2008}, $P_{\mathrm{Hom}} (1/2, d_0)~=~\frac{1}{2}\left[1-\mbox{Erf}(d_0/\sqrt{2})\right]$, as shown in  Fig.~\ref{fig:Helstrom}. 
The difference, $\Delta P$, between $P_{\mathrm{Hom}}$ and $P_{\mathrm{Hel}}$ characterizes the ``advantage'' of a quantum measurement over the classical one. This advantage is zero for $d_0 \to 0$ and $d_0 \to \infty$, and has a maximum at an intermediate separation.

\section{Discord potential} \label{sec:discordpot}

\begin{figure}[t]
\centering
    \includegraphics[scale=1]{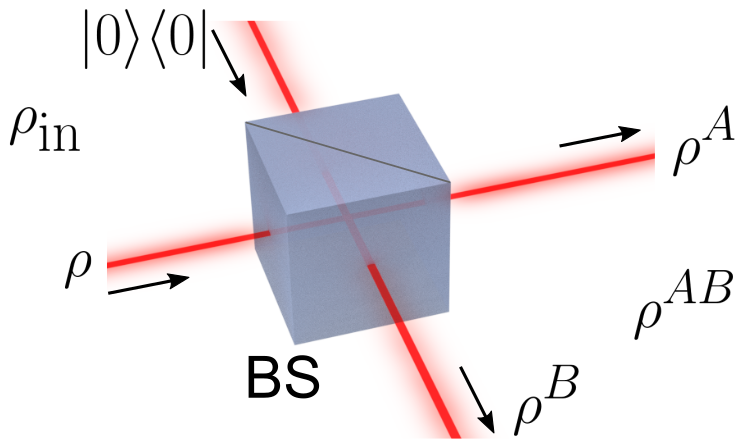}
    \caption{ \label{fig:ent} A beam splitter transforms a state $\rho$ and the vacuum state $| 0\rangle \langle 0|$ into a two-mode state $\rho^{AB}$. $\rho_A$ and $\rho_B$ are the reduced states of $\rho^{AB}$.}
\end{figure}

The advantage $\Delta P$ indicates a non-trivial quantum property in the state $\rho$ which we want to capture with a suitable measure of non-classicality. We note that the entanglement potential  was constructed as a measure of quantumness that explicitly identifies proper mixtures of coherent states as classical \cite{Asboth2005}; i.e. it does not capture this advantage.
Here we are looking for a measure $C$ that characterises the non-classicality of any state, with the following properties:
\begin{itemize}
\item it is positive-defined, 
\item it is non-zero for all states that have a non-zero entanglement potential,
\item it vanishes for the coherent state mixtures $\rho_0$ when $d_0 \to 0$ and when $d_0 \to \infty$,
\item and it is strictly positive for intermediate distances $d_0$ for mixtures $\rho_0$.
\end{itemize}
We define the \emph{discord potential $C_D$} as a measure of non-classicality of any state $\rho$ as
\begin{equation}
 	C_D (\rho) \equiv D (\rho^{AB}) ,
 \label{eq:criterion}
\end{equation}
where $D (\rho^{AB})$ is the discord of a two-mode state $\rho^{AB}$ obtained by impinging $\rho$ on a balanced beam splitter (BS), as shown in Fig.~\ref{fig:ent}. Formally, this output state can be written as 
\begin{equation}
\rho^{AB} = U_{BS} \left( \rho \otimes |0 \rangle \langle 0| \right) U_{BS}^{\dag},
\end{equation}
where $U_{BS}$ is the balanced BS unitary and $|0\rangle \langle0|$ is the vacuum state.

%%%
%%%

\medskip

The \emph{discord} $D_A(\sigma^{AB})$ of any two-mode state $\sigma^{AB}$ is a non-symmetrical correlation measure that quantifies the correlation between $A$ and $B$ when a von Neumann measurement $\{\Pi_j^A\}$, with orthogonal projectors $\Pi_j^A$, is performed on the mode $A$. It is defined as~\cite{Ollivier2001,Datta2008}:
\begin{equation}\label{eq:discord}
  	D_A(\sigma^{AB}) \equiv S(\sigma^{A}) - S(\sigma^{AB})+\min_{\{\Pi_j^A\}} S(\sigma^{B|\{\Pi_j^A\}}),
\end{equation}
where $ S(\sigma^{AB})$ is the von Neumann entropy of the global state and $S(\sigma^{A})$ is the entropy of the reduced state $\sigma^A$ of mode $A$. The last term involves the conditional entropy of the mode $B$ depending on the outcome of the measurement $\{\Pi_j^A\}$ on $A$, $S( \sigma^{ B|\{ \Pi_j^A \}}) = \sum_j p_j \, S(\tr_A \left[ \Pi_j^A \sigma^{AB} \Pi_j^A \right])$, where $p_j = \tr \left[ \Pi_j^A \sigma^{ AB} \Pi_j^A \right]$ is the probability for the outcome $\Pi_j^A$ to occur. 
Note, that the discord is defined as the \emph{minimum} over all possible von Neumann measurements, i.e. sets of projectors on $A$.

\medskip

As the BS output states $\rho^{AB}$ are symmetrical by construction, $D_A (\rho^{AB}) = D_B (\rho^{AB})$, which we denote by $D(\rho^{AB})$ in the definition of the discord potential $C_D$ in Eq.~(\ref{eq:criterion}). $C_D$ is also positive-definite by construction since $D$ is positive-definite. Furthermore, since the discord of any  entangled state $\sigma^{AB}$ is non-zero ~\cite{Ollivier2001,Datta2008}, $C_D$ is non-zero for all states with non-zero entanglement potential \cite{Asboth2005}.

\section{Properties of the discord potential for mixtures of two coherent states.}   \label{sec:properties}

We are now interested in the discord potential of mixtures of two coherent states $\rho_0$ for which we need to characterise the two-mode output state. One way to obtain $\rho^{AB}$ is to use well-known relations between the quasi-probability distributions of the input and output of a lossless balanced beam splitter~\cite{Ou1987}. 
Specifically, for a one-mode state $\rho$ the P-function is a distribution over a complex amplitude $\xi$,
\begin{equation}
	{\cal P}_{\rho} (\xi)=\frac{e^{|\xi|^2}}{\pi} \, \int \langle -\gamma| \, \rho \, |\gamma \rangle \, e^{|\gamma|^2- \gamma \xi^*+\gamma^*\xi} \, \mbox{d}^2 \gamma,
\end{equation}
where the integration is performed over all coherent states $|\gamma\rangle$ with complex amplitude $\gamma$.
Then the P-function of the two-mode output state is related to the P-function of the two-mode input state  as \cite{Loudon2000,Glauber1963}: 
\begin{equation} \label{eq:P_input_output}
	{\cal P}_{\rho^{AB}}(\xi^\prime, \zeta^\prime) = {\cal P}_{\rho_0 \otimes |0\rangle \langle 0|} \, \left(\frac{\xi^\prime - i \zeta^\prime}{\sqrt{2}}, \frac{\zeta^\prime - i \xi^\prime}{\sqrt{2}}\right).
\end{equation}
For the input state $\rho_0 \otimes |0\rangle \langle 0|$ the P-function is:
\begin{equation} \label{eq:P_in}
	{\cal P}_{\rho_0 \otimes |0\rangle \langle 0|} \, (\xi, \zeta)= \left[ a \, \delta^2(\xi-\alpha_0) + (1-a) \, \delta^2(\xi-\beta_0) \right] \cdot \delta^2(\zeta).
\end{equation}
Substituting~\eref{eq:P_in} into~\eref{eq:P_input_output} gives the output P-function, from which we obtain the output state:
\begin{equation}\label{eq:out}
\rho^{AB} =a \,  \vert \alpha \rangle  \langle  \alpha \vert \otimes \vert i\alpha \rangle  \langle i\alpha  \vert +(1-a) \, \vert \beta \rangle  \langle  \beta \vert \otimes \vert i\beta \rangle  \langle i\beta  \vert,
\end{equation}
where $\alpha = \alpha_0/\sqrt{2}$, $\beta = \beta_0/\sqrt{2}$. The reduced state $\rho^A$ of the mode $A$ is then:
\begin{equation}\label{eq:rhoA}
\rho^A  = a \vert \alpha \rangle  \langle  \alpha \vert + (1-a) \vert \beta \rangle \langle \beta \vert,
\end{equation}
with $\rho^B$ taking exactly the same form. 

%%%
\begin{figure}[b]
\centering
    \includegraphics[scale=1]{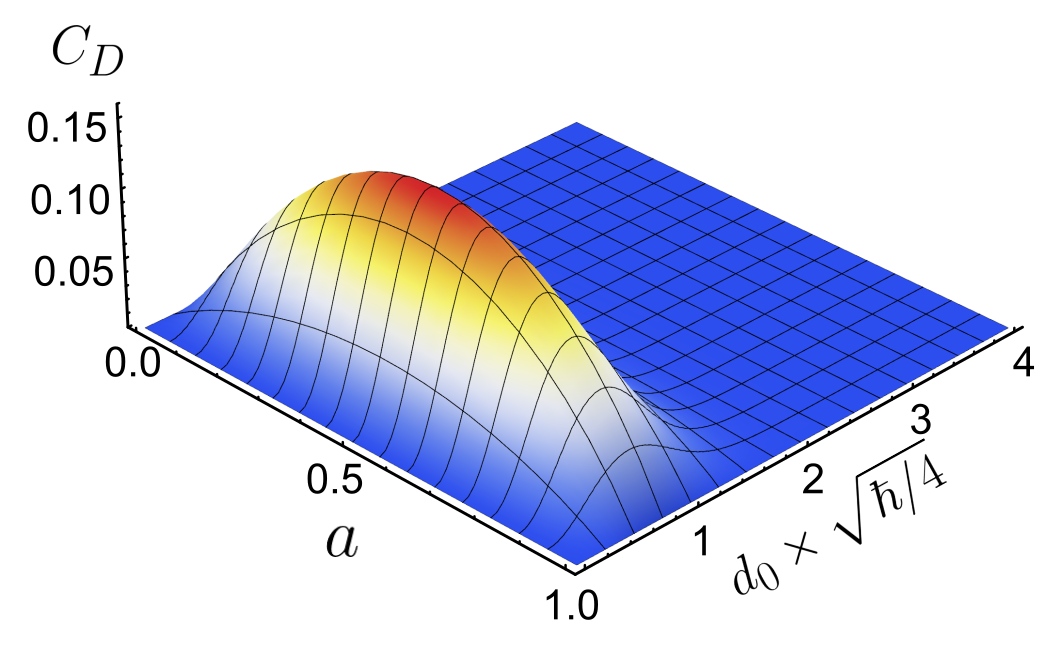}
    \caption{ \label{fig:reduced_ent} 
    Discord potential $C_D$ of $\rho_0$, see Eq.~\eref{eq:rho}, as a function of mixing probability $a$ and overlap $d_0$. }
\end{figure}
%%%

We are now ready to calculate  the discord $D (\rho^{AB})$. While calculations in the infinite dimensional Hilbert space can be challenging and are usually limited to Gaussian states~\cite{Giorda2010,Adesso2010,Giorda2012}, for the particular set of states considered here a straightforward method is discussed in Section \ref{sec:method}.  The discord of the two-mode output state, $D(\rho^{AB})$, and thus the discord potential $C_D (\rho_0)$ of the input state,  shown in Fig.~\ref{fig:reduced_ent}, vanishes for $d_0 \to 0$, $d_0 \to \infty$, $a \to 0$ and $a \to 1$, as we required for our non-classicality measure.

%%%
\begin{figure}[!t]
\centering
    \includegraphics[scale=1]{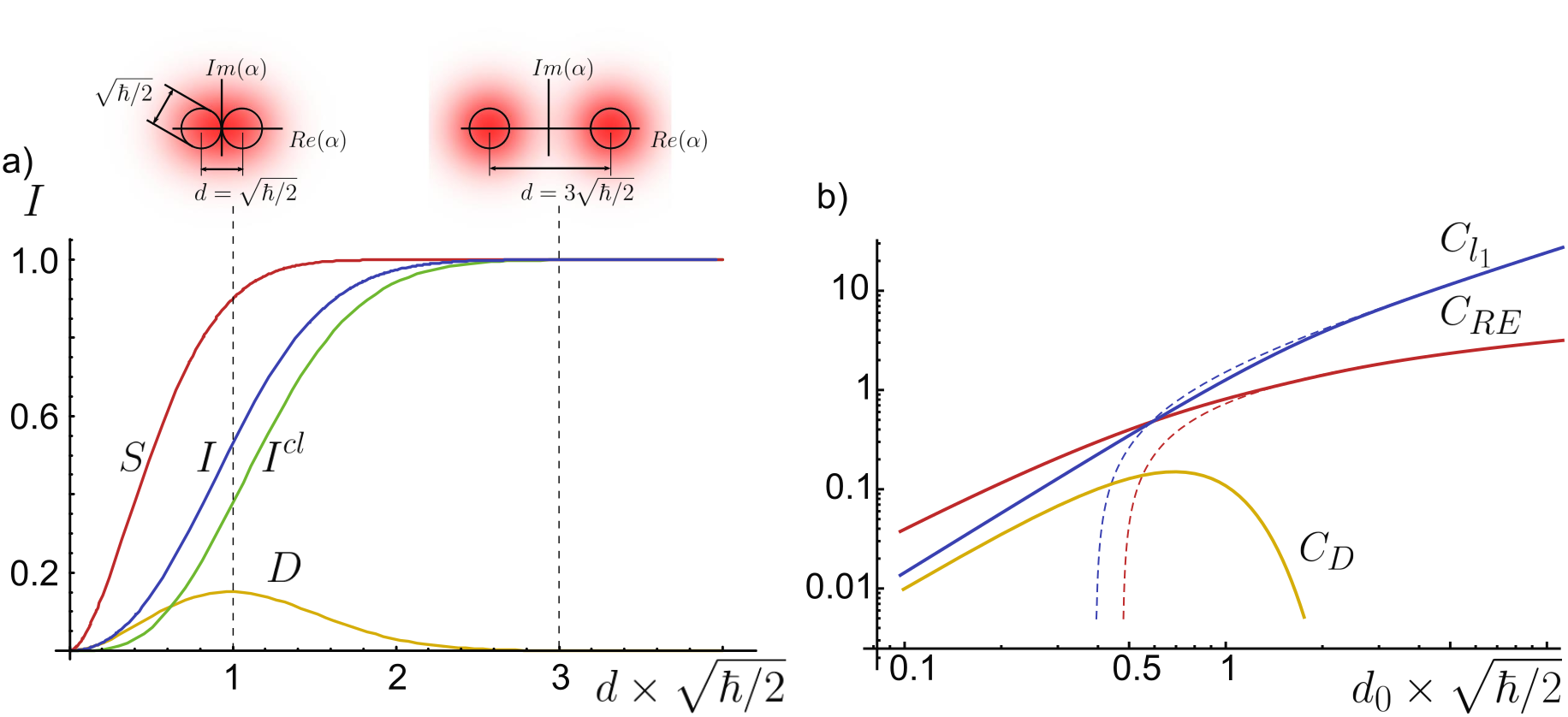}
    \caption{ \label{fig:discord} The input state is $\rho_0$ with ratio $a =1/2$ and $\beta_0 = - \alpha_0$. 
    {\bf a)} Output state's total entropy $S(\rho^{AB})$ (red line), its mutual information $I(\rho^{AB})$ (blue line), classical contribution to the mutual information $I^{cl}(\rho^{AB})$ (green line) and quantum contribution to the mutual information, i.e. discord, $D(\rho^{AB})$ (yellow line) as a function of $d$. 
The upper panel illustrates the overlap between the two coherent states $|\alpha\rangle$ and $|\beta\rangle$ in $\rho^{AB}$. 
   {\bf b)} Discord potential $C_D$ (solid yellow) and coherence monotones for $\rho_0$ as a function of $d_0=\sqrt{2} \, d$. The $l_1$ norm of coherence $C_{l_1}$ (solid blue) and  relative entropy of coherence $C_{RE}$ (solid red) are shown together with their asymptotic behaviour (dashed lines) for large $d_0$, see Appendix for detailed expressions.   
      }
\end{figure}
%%%

Fig.~\ref{fig:discord}a shows the output state discord $D(\rho^{AB})$, its entropy $S(\rho^{AB})$, its mutual information $I (\rho^{AB})=S(\rho^{A})+S(\rho^{B}) - S(\rho^{AB})$, and the ``classical'' contribution to the mutual information, $I^{cl}(\rho^{AB})=I (\rho^{AB}) - D (\rho^{AB})$, as a function of $d = |\alpha - \beta|$, with $d$ directly related to the separation in the input state, as $d= d_0/\sqrt{2}$. As can be seen from the figure, the discord reaches a maximum for $ d = \sqrt{\hbar/2}$. This is the distance at which the peaks of the input state Wigner-function is equal to half of the value of the coherent state quadrature uncertainty, which is equivalent to the Rayleigh criterion for peak resolution~\cite{Born1994}. Interestingly, for $d \lessapprox  \sqrt{\hbar/8}$ the quantum discord contribution to $I$ becomes larger than the classical contribution.

Alternative candidates that may satisfy our requirements  for a measure of non-classicality  are the coherence monotones~\cite{Baumgratz2014,Streltsov2016}, which quantify the coherence of a quantum state with respect to a chosen basis. In quantum information theory quantum coherences have been shown to be a resource for creating non-classical correlations~\cite{Streltsov2015,Killoran2016}, for enhancing quantum measurement precision~\cite{Marvian2016} and for performing certain quantum computational tasks~\cite{Hillery2016}. Coherences can be quantified by the \emph{$l_1$ norm of coherence}
\begin{equation}\label{eq:l1norm}
	C_{l_1}(\rho) = \sum_{n\ne k} | \rho_{n,k}|,
\end{equation}
where $\rho_{n,k}$ are the coefficients of the state in a basis $\{| n \rangle\}$, and the \emph{relative entropy of coherence}
\begin{equation}\label{eq:relentrcoh}
	C_{RE}(\rho) = S(\rho_{\mathrm{diag}}) - S(\rho),
\end{equation}
where $\rho_{\mathrm{diag}}$ is the state obtained by removing the non-diagonal elements of $\rho$ in the chosen basis. 

Choosing the energy basis (i.e. the Fock basis), these measures are plotted in Fig.~\ref{fig:discord}b for the mixture of two coherent states $\rho_0$ with $a=1/2$ and $\beta_0 = -\alpha_0$ as a function of separation $d_0$. One can see that both, the $l_1$ norm of coherence and the relative entropy of coherence,  increase monotonously with $d_0$. Notably they diverge for $d_0 \to \infty$, which contrasts with the properties we require for our non-classicality measure. Indeed, the $l_1$ norm of coherence has already been shown to diverge for a certain class of states  in the infinite-dimensional case~\cite{Zhang2016}. We here prove in the Appendix (Section \ref{sec:app}) that actually both measures diverge for the coherent state mixtures. Moreover, both coherence monotones depend not only on the separation between the coherent states in the mixture $d_0$, but they also depend on the absolute amplitude of each coherent state $|\alpha_0|$ and $|\beta_0|$. In contrast, $C_D(\rho)$ depends only on $d_0$, tends to zero at both small and large values of $d_0$, and quantifies the state's non-classicality at intermediate separations, as can be seen in Fig.~\ref{fig:discord}b. Thus the discord potential satisfies all our requirements. In addition, $C_D$ does not require to choose a basis, in contrast to the coherence monotones.

%%%%
\section{Calculating entropies of (non-Gaussian) coherent state mixtures}  \label{sec:method}

Coherent states, such as~\eref{eq:rho}, are elements of the infinite dimensional Hilbert space spanned by the Fock basis. At first this makes the direct calculation of the entropies in~\eref{eq:discord} tricky, as  one would need to find the eigenvalues of an infinite matrix. However, these quantities can be obtained  by moving to the Hilbert space spanned by the non-trivial pure state elements of the considered mixture and establishing an orthonormal basis in this smaller sub-space. 
For a general proper mixture of coherent states \begin{equation}\label{eq:rho2}
	\rho = \sum_{j=1}^N p_j \, |\alpha_j\rangle \langle \alpha_j |, 
\end{equation}
with $p_j>0$, this subspace can be spanned by the $| \alpha_j \rangle$ for $j = 1, \dots, N$, the orthonormal basis $|u_j\rangle$ can be built using the Gram-Schmidt procedure:
\begin{eqnarray} \label{eq:basis}
	|u_1\rangle& = | \alpha_1 \rangle , \\ 
	|u_j\rangle &= \frac{|v_j\rangle}{|| v_j ||}, \quad |v_j\rangle 
	&=  | \alpha_j \rangle - \sum_{k}^{N-1} \frac{\langle v_k| \alpha_j \rangle}{\langle v_k| v_k \rangle},
\end{eqnarray}
In such a basis the only non-zero elements of $\rho$ will be:
\begin{equation}\label{eq:rhomn}
	\rho_{jk}=\langle u_j | \rho | u_k \rangle.
\end{equation}
This is a finite dimensional matrix of size $N \times N$, where $N$ is the number of pure state elements in Eq.~\eref{eq:rho2}. This finite-dimensional matrix is straightforwardly diagonalized allowing the calculation of the eigenvalues and entropy of the state $\rho$.

\medskip

For example, for the reduced state $\rho^A$ in Eq.~\eref{eq:rhoA} an orthonormal basis in the subspace spanned by $|\alpha \rangle$ and $|\beta \rangle$ is:
\begin{equation}
	\vert u_1\rangle  = \vert \alpha \rangle; \qquad 
	\vert u_2 \rangle = \frac{\vert \beta \rangle - k \vert \alpha \rangle}{\sqrt{1-|k|^2}},
\end{equation}
where $k =\langle \alpha \vert \beta \rangle$. In this basis $\rho^A$ can be written as:
\begin{equation}
	\rho^{A}_\bot 
		= \left(
	\begin{array}{cc}
 	a + (1-a) k^2 
	& \frac{k(1-a)(1-|k|^2)}{\sqrt{1-k^2}} \\
	\frac{k^*(1-a)(1-|k|^2)}{\sqrt{1-k^{*2}}} & (1-a)(1 -k^2)
	\end{array}
		\right).
\end{equation}
The eigenvalues and the entropy, $S(\rho^A)$, are now readily calculated. 
To calculate the entropy of the two-mode state $S(\rho^{AB})$  in Eq.~\eref{eq:out} one has to introduce an analogous pair of basis vectors for the second mode and then diagonalise a four-dimensional matrix to obtain the entropy.
The last term in the expression for the discord \eref{eq:discord} is the entropy of a single reduced mode, but requires the optimization over all possible measurement operators for the other mode. In the subspace spanned by $|\alpha \rangle$ and $|\beta \rangle$ the general set of measurement operators is:
\begin{eqnarray}
	\Pi_1^A &= \cos \theta \, \vert u_1 \rangle + e^{i\phi} \, \sin \theta \, \vert u_2 \rangle \cr
	\Pi_2^A &= \sin \theta \, \vert u_1 \rangle - e^{-i\phi} \,  \cos \theta \, \vert u_2 \rangle
\end{eqnarray}
Using these operators it is then possible to calculate the conditional entropy, $S(\sigma^{B|\{\Pi_j^A\}})$ and minimize this entropy over all $\theta$ and $\phi$.

\medskip

For the more general case that the input state is a mixture of more than two pure state elements, i.e. $N>2$, the Gram-Schmidt diagonalization can still be used to calculate the entropies for  $N\otimes N$ reduced states. 
But the optimization procedure for obtaining the conditional entropies becomes complicated and has so far been shown to be possible using the linear entropy approximation for $2\otimes N$ systems~\cite{Ma2015}.

%%%
\section{Conclusions} \label{sec:concl}

Despite the existence of a huge variety of non-classicality criteria, it still can be hard to determine if a given system requires a full quantum-mechanical description or not, especially if it is infinite-dimensional. 
In particular, mixtures of coherent states are often considered classical, but there is a quantum advantage when discriminating between overlapping coherent states in the mixture, suggesting that a quantum-mechanical description is necessary to fully capture  the system properties.
We proposed a new criterion for non-classicality of a state $\rho$ based on the quantum discord of the output $\rho^{AB}$ of a balanced beam splitter, when the two inputs are the state of interest $\rho$ and the vacuum state $|0 \rangle \langle 0|$. This measure, the discord potential $C_D$, identifies as non-classical all states with non-zero entanglement potential, but it is also positive for mixtures of coherent states when the quantum advantage in discrimination is positive, whilst vanishing otherwise. We also showed that the discord potential has several advantages over another set of commonly used non-classicality measures: the coherence monotones. Namely, the discord potential does not require a basis choice and it does not diverge for mixtures between classically separated coherent states, in contrast to two coherence monotones. Furthermore, we detailed a simple method to calculate the entropy of any state with positive P-function, and showed how to calculate the quantum potential of mixtures of two coherent states requiring optimization over all possible measurements. 
We conclude that the discord potential can be a more sensitive indicator of non-classicality than the entanglement potential or the positivity of the P-function, capturing a wider class of states that show quantum advantages. 

\medskip

\section{Acknowledgements}
IS acknowledges support from EPSRC (EP/L015331/1) through the Centre of Doctoral Training in Metamaterials (XM$^2$).
JB acknowledges support from the Leverhulme Trust's Philip Leverhulme Prize.
JA acknowledges support by EPSRC (EP/M009165/1). 

%%%%
\section*{References}

\bibliographystyle{spphys}       % APS-like style for physics

%\begin{thebibliography}{}

\bibliography{references} 

\begin{thebibliography}{10}
\providecommand{\url}[1]{{#1}}
\providecommand{\urlprefix}{URL }
\expandafter\ifx\csname urlstyle\endcsname\relax
  \providecommand{\doi}[1]{DOI \discretionary{}{}{}#1}\else
  \providecommand{\doi}{DOI \discretionary{}{}{}\begingroup
  \urlstyle{rm}\Url}\fi

\bibitem{Ferraro2012}
A.~Ferraro, M.G. Paris, Physical Review Letters \textbf{108}(26), 260403 (2012)

\bibitem{Dodonov}
V.~Dodonov, V.~Man'ko, \emph{{Theory of non-classical states of light}} (1994)

\bibitem{Ryl2015}
S.~Ryl, J.~Sperling, E.~Agudelo, M.~Mraz, S.~K{\"{o}}hnke, B.~Hage, W.~Vogel,
  Physical Review A \textbf{92}, 011801 (2015)

\bibitem{Yuan2018}
X.~Yuan, H.~Zhou, M.~Gu, X.~Ma, Physical Review A \textbf{97}, 012331 (2018)

\bibitem{Bernardini2017}
A.E. Bernardini, O.~Bertolami, EPL (Europhysics Letters \textbf{120}, 20002
  (2017)

\bibitem{Gilchrist1997}
A.~Gilchrist, C.W. Gardiner, P.D. Drummond, Physical Review A \textbf{55}(4),
  3014 (1997)

\bibitem{Glauber1963}
R.J. Glauber, Physical Review \textbf{130}(6), 2529 (1963)

\bibitem{Asboth2005}
J.K. Asb{\'{o}}th, J.~Calsamiglia, H.~Ritsch, Physical Review Letters
  \textbf{94}(17), 173602 (2005)

\bibitem{Baumgratz2014}
T.~Baumgratz, M.~Cramer, M.B. Plenio, Physical Review Letters \textbf{113}(14),
  140401 (2014)

\bibitem{Streltsov2016}
A.~Streltsov, G.~Adesso, M.B. Plenio, Reviews of Modern Physics \textbf{89},
  041003 (2016)

\bibitem{Loudon2000}
R.~Loudon, \emph{{The Quantum Theory of Light}} (2000)

\bibitem{Helstrom1967}
C.W. Helstrom, Information and Control \textbf{10}(3), 254 (1967)

\bibitem{Barnett2009}
S.M. Barnett, S.~Croke, Advances in Optics and Photonics \textbf{1}(2), 238
  (2009)

\bibitem{Hosseini2014}
S.~Hosseini, S.~Rahimi-Keshari, J.Y. Haw, S.M. Assad, H.M. Chrzanowski,
  J.~Janousek, T.~Symul, T.C. Ralph, P.K. Lam, Journal of Physics B: Atomic,
  Molecular and Optical Physics \textbf{47}, 025503 (2014)

\bibitem{Choi2017}
Y.~Choi, K.H. Hong, H.T. Lim, J.~Yune, O.~Kwon, S.W. Han, K.~Oh, Y.H. Kim, Y.S.
  Kim, S.~Moon, Opt. Express \textbf{25}(3), 2540 (2017)

\bibitem{Wittmann2008}
C.~Wittmann, M.~Takeoka, K.N. Cassemiro, M.~Sasaki, G.~Leuchs, U.L. Andersen,
  Physical Review Letters \textbf{101}(21), 210501 (2008)

\bibitem{Cook2007}
R.L. Cook, P.J. Martin, J.M. Geremia, Nature \textbf{446}(7137), 774 (2007)

\bibitem{Becerra2013}
F.E. Becerra, J.~Fan, A.~Migdall, Nature Communications \textbf{4}, 2028 (2013)

\bibitem{Rosati2016}
M.~Rosati, A.~Mari, V.~Giovannetti, Physical Review A \textbf{93}, 062315
  (2016)

\bibitem{Ollivier2001}
H.~Ollivier, W.H. Zurek, Physical Review Letters \textbf{88}(1), 017901 (2001)

\bibitem{Datta2008}
A.~Datta, A.~Shaji, C.M. Caves, Physical Review Letters \textbf{100}(5), 050502
  (2008)

\bibitem{Ou1987}
Z.~Ou, C.~Hong, L.~Mandel, Optics Communications \textbf{63}(2), 118 (1987)

\bibitem{Giorda2010}
P.~Giorda, M.G.A. Paris, Physical Review Letters \textbf{105}(2), 020503 (2010)

\bibitem{Adesso2010}
G.~Adesso, A.~Datta, Physical Review Letters \textbf{105}(3), 030501 (2010)

\bibitem{Giorda2012}
P.~Giorda, M.~Allegra, M.G.A. Paris, Physical Review A \textbf{86}, 052328
  (2012)

\bibitem{Born1994}
M.~Born, E.~Wolf, \emph{{Principles of optics: Electromagnetic Theory of
  Propagation, Interference and Diffraction of Light}} (1994)

\bibitem{Streltsov2015}
A.~Streltsov, U.~Singh, H.S. Dhar, M.N. Bera, G.~Adesso, Physical Review
  Letters \textbf{115}(2), 020403 (2015)

\bibitem{Killoran2016}
N.~Killoran, F.E. Steinhoff, M.B. Plenio, Physical Review Letters
  \textbf{116}(8), 080402 (2016)

\bibitem{Marvian2016}
I.~Marvian, R.W. Spekkens, P.~Zanardi, Physical Review A \textbf{93}(5), 052331
  (2016)

\bibitem{Hillery2016}
M.~Hillery, Physical Review A \textbf{93}(1), 012111 (2016)

\bibitem{Zhang2016}
Y.R. Zhang, L.H. Shao, Y.~Li, H.~Fan, Physical Review A \textbf{93}(1), 012334
  (2016)

\bibitem{Ma2015}
Z.~Ma, Z.~Chen, F.F. Fanchini, S.M. Fei, Scientific Reports \textbf{5}, 10262
  (2015)

\end{thebibliography}
%\end{thebibliography}

%%%%%
%%%%%
%%%%%

\section{Appendix} \label{sec:app}

\subsection{Asymptotic behaviour of $C_{l_1}$ for coherent states in the Fock basis.} 

We first evaluate the large $\alpha$ asymptotes of the coherence monotone $C_{l_1}$ for a coherent state $|\alpha\rangle$ in the Fock basis $\{|n \rangle \}$ with $n = 0, 1, 2, ....$, and then conclude with stating the asymptotic behaviour of a mixture of two coherent states.

Using Eq.~\eref{eq:l1norm} the $l_1$ norm of coherence for a coherent state in the Fock basis is:
\begin{eqnarray}
	C_{l_1}(|\alpha\rangle) 
	&= e^{-|\alpha|^2} \, \sum_{k \ne n} \frac{|\alpha|^{n+k}}{\sqrt{k!\, n!}}
	= e^{-|\alpha|^2} \, \left( \left[1 + \sum_{k=1}^\infty \frac{|\alpha|^{k}}{\sqrt{k!}} \right]^2 - \sum_{k=0}^\infty \frac{|\alpha|^{2k}}{k!} \right).
\end{eqnarray}

The resulting summation contains a square root of a factorial, and does not have a closed form. However, its asymptotic behaviour at large $A:=|\alpha|$ can be estimated using the following procedure: 

We introduce
\begin{equation}
	f(A) =\sum_{k=1}^\infty \frac{A^k }{\sqrt{k!}} 
	\quad \mathrm{and} \quad 
	g(A) = {A \over f (A)} \, \left({\mbox{d} f \over \mbox{d} A}\right),
\end{equation}
so that
\begin{eqnarray}
	C_{l_1}(|\alpha\rangle) 
	&= e^{- A^2} \, \left(1 + 2 f(A) + f^2(A) \right) - 1.
\end{eqnarray}
Using that the ratio of two power series in $x$ with coefficients $p_k$ and $q_k$ for large $x$ is asymptotically determined by the ratio of the coefficients of the largest power,
\begin{equation}
	\lim_{x\to \infty} \frac{\sum_{k=1}^\infty p_k \, x^k}{\sum_{k=1}^\infty q_k \, x^k} 
	=\lim_{k\to \infty} \frac{p_k}{q_k},
\end{equation}
the asymptote of $g(A)/A^2$ for large $A$ is:
\begin{equation}
	\lim_{A \to \infty} \frac{g(A)}{A^2} 
	=  \lim_{A \to \infty}  \frac{\sum_{k=0}^\infty \frac{(k+1) A^{k} }{\sqrt{(k+1)!}}}{\sum_{k=2}^\infty \frac{A^{k} }{\sqrt{(k-1)!}} } 
	=\lim_{k\to \infty} {(k+1)  \sqrt{(k-1)!} \over \sqrt{(k+1)!}} = 1,
\end{equation}
and thus $g(A) = A^2 + o(A^2)$. The next order of the approximation is given by 
\begin{eqnarray}
	\lim_{A \to \infty} \left(g(A) - A^2\right) 
	=  \lim_{A \to \infty}  \frac{\sum_{k=2}^\infty \frac{(k-1) A^{k} }{\sqrt{(k-1)!}} - \sum_{k=4}^\infty \frac{A^{k} }{\sqrt{(k-3)!}}}{\sum_{k=2}^\infty \frac{A^{k} }{\sqrt{(k-1)!}} } \\ 
	=\lim_{k\to \infty} \left((k-1) - {\sqrt{(k-1)(k-2)  }}\right) = \frac{1}{2},
\end{eqnarray}
and thus $g(A) = A^2 + \frac{1}{2}+o(1)$. 

This leads to the following differential equation for $f(A)$:
\begin{equation}
	\left({\mbox{d} f \over \mbox{d} A}\right) = \left( A^2 + \frac{1}{2}+o(1) \right) \, {f(A) \over A},
\end{equation}
which in the asymptotic limit of large $A$ has the solution
\begin{equation}
	f (A) \approx e^{A^2 \over 2} \, \sqrt{c A}
\end{equation} 
with $c$ a constant which can be evaluated numerically, resulting in $c \approx 5$. Further terms of the approximation will lead to multipliers of the form $\exp[{1 \over A^k}]$ with $k>1$ in $f(A)$, which quickly converge to 1 for large $A$.
This allows one to conclude that the $l_1$ norm of coherence of a coherent state in the Fock basis for large $\alpha$ asymptotically becomes
\begin{eqnarray}
	C_{l_1}(|\alpha\rangle) 
	&= e^{- |\alpha|^2}  + 2 e^{- |\alpha|^2 \over 2} \, \sqrt{c |\alpha|} + c |\alpha|  - 1 \approx c |\alpha|
\end{eqnarray}
i.e. that $C_{l_1}$ diverges for $|\alpha| \to \infty$. 

We now return to a proper mixture of two coherent states, $\rho  = a \vert \alpha \rangle  \langle  \alpha \vert + (1-a) \vert \beta \rangle \langle \beta \vert$ with $0 < a <1$. This state's coefficients in the Fock basis are 
\begin{equation}
	\rho_{k,n} = \frac{a \, e^{-|\alpha|^2} \alpha^k\alpha^{*n}+(1-a)e^{-|\beta|^2} \beta^k\beta^{*n}}{\sqrt{k! \, n!}}.
\end{equation}
One can see that these coefficients imply that the $l_1$ norm of coherence, Eq.~\eref{eq:l1norm}, will depend on the absolute values of $\alpha$ and $\beta$, not just their relative displacement, in contrast to the discord potential which only depends on the relative displacement. 

To illustrate the behaviour of the $l_1$ norm of coherence on the separation of the elements of the mixture we here choose  $\beta$ = $-\alpha$ and also $a=1/2$. The mixed state coefficients then simplify to
\begin{equation}
	\rho_{k,n} =  \frac{\alpha^k\alpha^{*n} }{\sqrt{k! \, n!}} \,  e^{-|\alpha|^2} \quad \mbox{for} \quad k+n = \mbox{even}.
\end{equation}
Thus, the coefficients of the mixture of coherent states are identical to those for the coherent state $| \alpha \rangle$, but only when $k+n$ is an even number. As a consequence the coherence monotone $C_{l_1}(\rho)$ has almost identical asymptotic behaviour for large $|\alpha|$ as the monotone for the coherent state $C_{l_1}(|\alpha\rangle)$, \green{with $C_{l_1}(\rho) = \frac{1}{2} \, C_{l_1}(|\alpha\rangle) \approx \frac{c}{2} |\alpha| $.}  The diverging asymptotic behaviour of $C_{l_1} (\rho)$ for large $\alpha$ is indicated in Fig.~\ref{fig:discord}.

%%%
\subsection{Asymptotic behaviour of $C_{RE}$ for coherent states in the Fock basis.} 

We first evaluate the large $\alpha$ asymptotes of the relative entropy of coherence $C_{RE}$ for a coherent state $|\alpha\rangle$ in the Fock basis $\{|n \rangle \}$ with $n = 0, 1, 2, ....$, and then conclude with stating the asymptotic behaviour of a mixture of two coherent states.

The second term in Eq.~\eref{eq:relentrcoh} is 0 since the coherent state is pure, which leaves us with
\begin{eqnarray}
	C_{RE}(|\alpha\rangle) &= - e^{-|\alpha|^2}  \sum_{k =0}^\infty \frac{|\alpha|^{2k}}{k!}\ln \left(e^{-|\alpha|^2} \frac{|\alpha|^{2k}}{k!} \right)\\
&=|\alpha|^2 (1- 2\ln |\alpha|) + e^{-|\alpha|^2} \sum_{k =0}^\infty \frac{|\alpha|^{2k}}{k!}\ln k! .
\end{eqnarray}
For large $A:=|\alpha|$ the high powers are important and we use the Stirling approximation $\ln k! \approx k \ln k - k +\frac{1}{2} \ln (2\pi k) + \dots$, to transform the above expression to:
\begin{eqnarray}
\fl	C_{RE}(|\alpha\rangle) 
	&=  \frac{1}{2}e^{-A^2}  \sum_{k =1}^\infty \frac{A^{2k}}{k!}\ln k  
	+A^2 e^{-A^2}  \sum_{k =0}^\infty \frac{A^{2k}}{k!}\ln (k+1)- A^2 \ln A^2 +\frac{\ln(2\pi)}{2}.
\end{eqnarray}
Now we need to establish the asymptotic behaviour  for large $A$ of the functions
\begin{equation}\label{eq:h0}
	h_0(A) =e^{-A^2} \sum_{k =1}^\infty \frac{A^{2k}}{k!}\ln k \quad \mbox{and} \quad 
	h_1(A) =e^{-A^2} \sum_{k =0}^\infty \frac{A^{2k}}{k!}\ln(k+1).
\end{equation}
After substituting the Laplace transform identity $\ln k = - k \int_0^\infty e^{-kt}\ln t \,  \mbox{d} t -\gamma$, in Eq.~\eref{eq:h0}, where $\gamma$ is the Euler -- Mascheroni constant, $h_0(A)$ becomes:
\begin{eqnarray}
	h_0(A) &= -e^{-A^2} \int_0^\infty \sum_{k =1}^\infty \frac{A^{2k} e^{-kt}}{k!} k\ln t \,  \mbox{d} t -\gamma \, (1 - e^{-A^2})\\ 
	 &=  - e^{-A^2} A^2\int_0^\infty e^{-t+A^2 e^{-t}}\ln t \, \mbox{d} t -\gamma \, (1 - e^{-A^2})\\
	 &= -  A^2 \int_0^1 e^{A^2 (x-1)} \, \ln \left( \ln {1 \over x} \right) \, \mbox{d} x -\gamma  \, (1 - e^{-A^2})
\end{eqnarray}
% Mathematica - checked
where $x = e^{-t}$ and the integral kernel is $I(A,x) = e^{A^2 (x-1)} \, \ln \left( \ln {1 \over x} \right)$. 

The kernel $I(A,x)$ diverges at the points $x=0$ and $x=1$, i.e. $I(A,0) \to \infty$ and $I(A,1) \to -\infty$, and these points will give maximal contribution to the integral. Also at $x=1/e$ the kernel vanishes, i.e. $I(A, 1/e) = 0$. Splitting the integral into two parts, 
\begin{equation}\label{eq:I}
	\int_0^1 I(A,x) \, \mbox{d}  x  =  \int_0^{1/e} I(A,x) \, \mbox{d}  x  + \int_{1/e}^1 I(A,x) \, \mbox{d}  x \, ,
\end{equation}
we bound the asymptotic behaviour of the first integral using the Cauchy--Schwarz inequality
\begin{equation}
  	0 \le \int_0^{1/e} I(A,x) \, \mbox{d}  x \, \le  \, \frac{\mathrm{const.}}{\sqrt{2} A} \, e^{- A^2\left(1 - \frac{1}{e}\right) },
\end{equation}
% Mathematica - checked
where $\mathrm{const.}$ is a number arising from integrating $\left(\ln(-\ln x)\right)^2$ and taking the square root. This shows that the first integral decays exponentially with $A \to \infty$. 

In order to find the asymptotic behaviour of the second integral we change variables $x - 1 = -  p$,
\begin{equation}
 	\int_{1/e}^1 I(A,x) \, \mbox{d} x = \int_0^{1-1/e} e^{-p A^{2} } \ln (-\ln(1-p))\, \mbox{d} p.
\end{equation}
Now we can expand the function $\ln(-\ln(1-p))$ around the point $p=0$,
\begin{eqnarray}
	 \int_{1/e}^1 I(A,x) \, \mbox{d} x 
	 &= \int_0^{1-1/e} e^{- p A^{2}} \left( \ln p +\sum_{k=1}^\infty \frac{(-1)^{n+1}}{n}  \left[ \sum_{k=2}^\infty \frac{p^{k-1}}{k}\right]^n \, \right) \, \mbox{d} p\\
&= \int_0^{1-1/e} e^{- p A^{2}}  \left( \ln p + \frac{p}{2}+\frac{5 p^2}{24}+\frac{90 p^3}{720}+\dots \right) \, \mbox{d} p \\
&= \int_0^{1-1/e} e^{- p A^{2}} \ln p  \, \mbox{d} p +  \sum_{j=1} \lambda_j \int_0^{1-1/e} e^{- p A^{2}}  p^j \, \mbox{d} p \, .
\end{eqnarray}
where $\lambda_j$ are the expansion coefficient of the logarithm for powers of $p$.
The leading contribution to the first of these integrals is
\begin{equation}
	\int_0^{1-1/e} e^{- p A^{2}}  \ln p \, \mbox{d} p  
	\approx - \frac{\gamma + \ln A^2}{A^2}, 
\end{equation}
% Mathematica - checked
plus some exponentially decaying terms in $A$. The second integral gives
\begin{equation}
\fl	 \sum_{j=1} \lambda_j \int_0^{1-1/e} e^{- p A^{2}}  \, p^j \, \mbox{d} p 
	=\sum_{j=1} \lambda_j \frac{\Gamma(1+j) - \Gamma(1+j, A^2 (1- \frac{1}{e}))}{A^{2(1+j)}}
	= O\left(\frac{\mathrm{const}}{A^4}\right),
\end{equation}
% Mathematica - checked
where $\Gamma(k)$ and $\Gamma(k, A)$ are the Gamma and incomplete Gamma functions, respectively.
Therefore for large $A$, when dropping decaying terms, one finds
\begin{eqnarray}
	h_0(A)  &= -  A^2 \int_0^1 I(A,x) \, \mbox{d} x -\gamma  \, (1 - e^{-A^2}) \\
	&\approx -  A^2 \left(- \frac{\gamma + \ln A^2}{A^2} \right) +  O\left(\frac{\mathrm{const}}{A^4}\right) -\gamma  
	\approx  \ln A^2. 
\end{eqnarray}
Using similar arguments one finds
\begin{equation}
	h_1(A) \approx \ln A^2 +\frac{1}{2 A^2}+o\left(\frac{\mathrm{const}}{A^2}\right).
\end{equation}
Finally, the asymptotic behaviour of $C_{RE}$ of a coherent state for large $|\alpha|=A$ is 
\begin{eqnarray}
	C_{RE}(|\alpha\rangle) 
	&=  \frac{1}{2} \, h_0(A) +A^2 \, h_1(A) - A^2 \ln A^2 +\frac{\ln(2\pi)}{2} \\
	&\approx \ln A +\frac{1}{2} +\frac{\ln(2\pi)}{2} +o\left(\mathrm{const}\right),
\end{eqnarray}
which diverges logarithmically as $A \to \infty$.

We return again to the mixture of two coherent states $|\alpha \rangle$ and $|- \alpha \rangle$, $\rho=\frac{1}{2}\vert \alpha \rangle  \langle  \alpha \vert +  \frac{1}{2} \vert - \alpha \rangle \langle - \alpha \vert$ which allows us to illustrate the behaviour of the relative entropy of coherence $C_{RE}$ on the separation $d=2 |\alpha|$. Since the diagonal coefficients of the mixed state, $\rho_{n,n}$, are identical to those for the coherent state $| \alpha \rangle$ its coherence monotone $C_{RE}(\rho)$ is identical to the coherent state one $C_{RE}(|\alpha\rangle)$ apart from the fact that the entropy $S(\rho)$ is now non-zero and rises to $\ln 2$ for large $|\alpha|$. Hence $C_{RE}(\rho) \approx C_{RE}(|\alpha\rangle) - \ln 2 \approx \ln |\alpha| $ for large $|\alpha|$ and this diverging asymptotic behaviour of $C_{RE} (\rho)$ is indicated in Fig.~\ref{fig:discord}.

\end{document}